\documentclass[manuscript,screen]{acmart}

\AtBeginDocument{%
  \providecommand\BibTeX{{%
    \normalfont B\kern-0.5em{\scshape i\kern-0.25em b}\kern-0.8em\TeX}}}

\setcopyright{acmlicensed}
\copyrightyear{2018}
\acmYear{2018}
\acmDOI{XXXXXXX.XXXXXXX}

\acmConference[Conference acronym 'XX]{Make sure to enter the correct
  conference title from your rights confirmation emai}{June 03--05,
  2018}{Woodstock, NY}
\acmISBN{978-1-4503-XXXX-X/18/06}

\usepackage{subfiles}
\usepackage{graphicx}
\usepackage{caption}
\usepackage{subcaption}
\usepackage{multirow}
\usepackage{xspace}
\usepackage{algorithm}
\usepackage{algorithmic}
\usepackage{amsmath}
\usepackage{mathtools}
\usepackage[skip=2pt]{caption}
\usepackage[skip=2pt]{subcaption}
\usepackage{balance}  
\usepackage{hyperref}
\usepackage{amsmath}
\usepackage[export]{adjustbox}
\usepackage{tikz}

\usepackage{makecell}
\usepackage{color}

\usepackage{wrapfig}

\newcommand{\out}[1]{{#1}}

\newcommand{\new}[1]{\out{{}}}
\newcommand{\news}[1]{\out{{}}}

\newcommand{\usersaid}[1]{\textit{#1}}
\newcommand{\sys}{{Handy}{}}
\usepackage[nolist,nohyperlinks,smaller]{acronym}

\renewenvironment{quote}{\begin{quote}\itshape}{\end{quote}}
\renewenvironment{quote}
  {\list{}{\rightmargin\leftmargin\itshape}%
   \item\relax}
  {\endlist}

\begin{acronym}
    \acro{AR}{Augmented Reality}
    \acro{GPS}{Global Positioning System}
    \acro{VR}{Virtual Reality}
    \acro{CV}{Computer Vision}
    \acro{CSCW}{computer-supported cooperative work}
    \acro{SME}{subject-matter expert}
    \acro{HMD}{head-mounted display}
\end{acronym}

\begin{document}

\title[Investigating Remote Hands-On Assistance for Collaborative Development of Embedded Systems]{Investigating Remote Hands-On Assistance for Collaborative Development of Embedded Systems}

\author{Yan Chen}
\email{ych@vt.edu}
\affiliation{%
  \institution{Department of Computer Science, Virginia Tech}
  \city{Blacksburg}
  \state{VA}
  \country{USA}
}
\author{Jasmine Jones}
\affiliation{
  \institution{Berea College, Berea, KY, USA}
  \city{Berea}
  \state{KY}
  \country{USA}
 }
\email{jonesj2@berea.edu}
\renewcommand{\shortauthors}{Chen and Jones}

\begin{abstract}
Developing embedded systems is a complex endeavor that frequently requires collaborative teamwork. With the rise of freelance work and the global shift towards remote work, the need for effective remote collaboration has become crucial for many developers and their clients. However, current communication and coordination tools are predominantly tailored for software development rather than hardware-focused tasks. This study investigates the potential for remote support tools specifically designed for embedded systems development. Through interviews with 12 experienced embedded systems developers, we explored their existing remote work practices, challenges, and requirements. We also conducted a user enactment study featuring a custom-designed remote manipulation agent, \sys{}, as a theoretical assistant, to identify the kinds of support developers would value in a collaborative setting. Our findings highlight the scenarios and strategies employed in remote work, the specific support needs, and the challenges related to information exchange, coordination, and execution. Additionally, we explore concerns around privacy, control, and trust when using remote physical manipulation tools. This research contributes to the field by integrating the development of embedded systems with the remote, on-demand collaboration and assistance typical of software environments, offering a solid empirical foundation for future research on remote manipulation agents in this area.
\end{abstract}




\maketitle

\section{Introduction}
Embedded systems development is integral to the field of physical computing, which integrates software with computational devices—such as microcontrollers, sensors, and actuators—that connect to and conduct actions in the physical world \cite{o2004physical}.
It is a complex task that requires knowledge and skill sets in both programming and physical computing \cite{booth2016crossed}.
Developers often rely on support from various external resources to do their work.
Search engines and community question-answering (CQA) websites (such as StackOverflow \cite{stack_overflow}) are popular resources for developers; however, studies have shown that other developers often provide the best support \cite{chen2016towards,herbsleb1995object,latoza2006maintaining}.
In contrast to web-based resources, expert developers can provide personalized assistance, high-level advice, and project-specific code. In addition, they can often help identify and correct bugs that an individual developer might miss on their own \cite{cathedral}.

Collaborating with other people over distance (i.e., remote and asynchronous aid, with freelancer support) has become more common and accessible in computing work and research in recent years. While existing tools allow developers to remotely share and collaborate on their code and research, many of them were designed to support software development rather than hardware development \cite{chen2017codeon, oney2018creating,park2018post}. As a result, current remote collaboration tools and strategies fall short of supporting the types of \textit{hands-on} tasks necessary for collaboration in physical computing \cite{karchemsky2019heimdall}.

Recently, due to advances in utility and cost reductions, researchers have begun investigating the use of teleoperated robot arms as desktop-scale tools to support direct manipulation in remote collaboration scenarios. A direct inspiration for the work presented in this paper is the Heimdall system for remote debugging \cite{karchemsky2019heimdall}.
Heimdall showed that a teacher could use customized hardware to perform hands-on tasks remotely, such as inspecting a student's circuits.
Furthermore, teleoperated robots are already used for complex tasks such as remote surgery \cite{avellino2020multimodal}.
{While Heimdall showed that advanced tools provide an opportunity for remote learning support, we believe a broader space remains unexplored.}
To begin exploring this design space, we conducted a two-phase study with 12 experienced embedded systems developers. First, we interviewed them regarding developers' current remote work practices, issues, and needs.  
\begin{itemize}
\item In what scenarios and for what purposes do embedded systems developers seek out remote support?
\item What challenges do embedded systems developers face when trying to work together remotely?
\end{itemize}

{To further explore the potential design space, }we then conducted a user enactment study \cite{odom2012enactment} envisioning developers' interactions with a customized remote manipulation tool, \sys{}, teleoperated by a hypothetical assistant, to elicit the range of supportive behavior that developers desire from a collaborator. In this study, we asked:
{
\begin{itemize}
\item What features would experienced developers envision in a remote collaboration tool to enable on-demand assistance on an embedded systems project?
\end{itemize}
}

Our findings describe the \textit{support and collaboration} scenarios in which remote work takes place; the information, coordination, and implementation needs expressed by embedded systems developers; and the privacy, control, and trust concerns that arise in the use of hands-on remote collaboration tools.
This research contributes to the literature by bringing remote, on-demand collaboration and help-seeking technology for embedded systems development in line with the more robust examples that already exist for software environments.
The empirical basis of this work provides a rich foundation of expressed needs, preferences, and desires to ground future work on remote manipulation tools and enhance collaboration support.

\section{Background and Related Work}
{Our work relates to three main areas: embedded system development support, on-demand help-seeking, and remote collaboration on physical tasks. In this section, we reveal gaps in the literature in these areas and the inform our research questions and study designs.}
\subsection{Providing Support to Embedded Systems Developers}

Much of the work in the physical computing field has investigated methods of reducing the learning curve for embedded system development. For example, one widely adopted platform is Arduino,\cite{banzi2008getting} which provides a low-cost and easy way for users of all expertise levels to develop embedded systems that interact with the physical environment. However, embedded system development is still a complex task that requires keen logic and skills, which can make it a challenge for both novice and experienced developers. For example, after conducting a study of a physical computing task, Booth et al. found that even experienced Arduino users had difficulties with starter projects\cite{booth2016crossed}.

Another thread of work has focused on modularizing the functions of the electronics and abstracting the circuit design to higher-level logic connections. For example, `Programmable Bricks'\cite{resnick1996programmable} used a LEGO brick assembly analogy to make it easier for children to rapidly prototype and program physical computing devices, and platforms such as Makerweaver\cite{kazemitabaar2017makerwear} and Microsoft .NET Gadgeteer~\cite{villar2010prototyping} have built on this concept. Other work has considered methods of easing the programming aspect, including approaches that use simplified block-based user interfaces, such as MakeCode~\cite{devine2018makecode} and other visual programming environments for physical computing platforms~\cite{hartmann2006reflective,millner2011modkit}). A third strand of support tools helps developers via learning environments like SHERLOCK~\cite{lesgold1988sherlock}, which teaches sophisticated electronics troubleshooting to fighter airplane engineers, or tools such as Fritzing~\cite{knorig2009fritzing}, which allows users to lay out circuits on a virtual breadboard to plan and share projects.

These tools support {learning} to develop embedded systems, and most are designed for beginners. Experienced developers also need support in implementing novel uses for common hardware, building custom hardware, and debugging challenging circuits. However, limited attention has been paid to the problems end-user developers face in physical computing tasks when designing support tools. Our work explores support that facilitates the work of more experienced developers.

Last but not least, Augmented Reality (AR) offers significant adaptability and customization for specific users and their unique needs in PCB design and development. Systems like ARDW facilitate a connection between virtual design files and the actual PCB through the use of projected AR and tracking of test probes~\cite{chatterjee2022ardw}. Augmented Silkscreen introduced several AR interaction methods aimed at aiding electrical engineers in PCB debugging, helping to reduce the frequent need to switch contexts~\cite{chatterjee2021augmented}. Our research further investigates user requirements in this domain by gathering insights from settings where remote assistance is sought.

\subsection{On-Demand Help Seeking}

Our work is closely related to the topic of on-demand help seeking~\cite{chen2020demand}. Since Web2.0, people have started to seek on-demand support not only from internal helpers within teams but also from external helpers, such as software engineering~\cite{chen2020bashon, chen2020improving}. For example, on-demand services on applications such as Uber and Task Rabbit match real-world tasks with helpers, offering streamlined courses of action and increasing market efficiency for microtasks~\cite{teodoro2014motivations}. Meanwhile, many community question-answering (CQA) websites have provided crowd expert support in the virtual world.

In the context of embedded systems development, on-demand help-seeking occurs when a requester creates a programming task and sends it to helpers. These helpers can be either team members or crowd workers, the latter being our target group. Developers may also seek crowd support from CQA websites such as Stack Overflow, but it often takes too long to get responses (over 11 minutes~\cite{mamykina2011design}). Unlike AnswerGarden~\cite{ackerman1990answergarden} or EdCode~\cite{chen2020edcode} where the help-seeking is within an organization or a learning session (meaning less context switching and less need to communicate context), we explore the potential for a more general-purpose support model in which helpers do not require prior familiarity with the task or with requesters. This makes capturing and presenting the task context challenging in terms of effective communication between requesters and helpers. Chen et al. conducted a related study to explore issues in communicating during on-demand help seeking in software development~\cite{chen2016towards}. They identified seven types of requests that developers make to receive help from the assistant: 1) memory aids, 2) explanatory requests, 3) high-level strategic guidance, 4) code requests, 5) bug fixing, 6) code refactoring, and 7) effort-saving requests. Following this line of work, we aim to explore new challenges and opportunities for real-time, on-demand assistance for embedded systems development.

\subsection{Remote Collaboration on Physical Tasks}

Our work builds on lessons learned from HCI on remote collaboration on physical tasks~\cite{fussell2000coordination, kraut1996collaboration}. Related literature has explored providing remote support on mechanical tasks such as construction with LEGO blocks~\cite{huang2013handsin3d,gao2016oriented,gao2017static,huang2013handsinair,kim2013comparing,gurevich2015design}, folding origami~\cite{fakourfar2016stabilized}, car repair~\cite{gauglitz2014world}, bike repair~\cite{kraut2003visual}, and learning physical tasks~\cite{li2022asteroids}. When collaborating on these physical tasks, audio and visual information are the most important information to exchange. Tools that provide audio and visual information to long-distance collaborators have been available for decades. Many prior studies have investigated how these techniques affect the efficacy and manner of communication and coordination efforts. For example, early work on remote collaboration using these tools has shown that audio-only communication is not as efficient as audio/video communication~\cite{fussell2000coordination}. However, audio/video communication is still less effective than co-located communication. Researchers have argued that depending on the task, collaborators may require different types of visual information to facilitate communication~\cite{fussell2000coordination}.

\subsubsection{Remote Collaboration on Embedded Systems Development}

In the context of embedded system development, remote collaboration becomes even harder. The information required for reasoning about mechanical tasks is often immediately visible: for example, with a furniture assembly task, each component and the progress toward the goal is obvious at a glance. However, with embedded systems, not all information is visible, such as a component's current or the code embedded into a chip. One prior study found that some developers working on embedded systems projects incorrectly believed an issue was the fault of the program rather than the circuit, leading to further issues~\cite{booth2016crossed}.

While the AR techniques we discussed show promise, they were designed with mechanical tasks in mind. To adapt these techniques to embedded systems development, recent tools such as Bifröst facilitate more efficient debugging by capturing and displaying development data in real time\cite{mcgrath2017bifrost}. Additionally, Heimdall allows instructors to remotely inspect, measure, and rewire students’ circuits~\cite{karchemsky2019heimdall}. However, both tools have drawbacks: use of Bifröst is limited to specific tasks, and developers must still rely on their own capacity to complete their tasks. Heimdall requires students to build their circuits on a specialized workbench, limiting the size of students’ projects and prohibiting them from working in parallel~\cite{karchemsky2019heimdall}. Our work explores the design and use of remote support tools that can scale assistance for embedded systems development tasks.

\section{Study Design}
We conducted a two-phase study to investigate current and desired practices in remote collaboration: a semi-structured interview and a lab-based technology probe study. Each phase involved 12 embedded systems developers who had remote collaboration experience (Table~\ref{tab:study3_stat}). We recruited participants by reaching out to local makerspaces, embedded systems research labs, and online gig-work markets frequented by those working on electronics and hardware projects (e.g., Upwork, Instructables). Participants reported two occupations: professional developers paid to develop embedded systems projects (n=9) and student developers who identified themselves as hobbyists, makers, or researchers of embedded systems development (n=2). Each participant self-identified as an experienced developer and was currently working on a relevant project with at least one other person. We will use P1-P12 to represent participants. All study interactions were conducted via Zoom.

\subsection{Phase 1: Semi-structured Interviews}
Interviews were semi-structured with questions about participants' work practices, collaboration strategies, and support challenges. We also asked participants to describe their preferred collaboration tools and resources. Each interview was conducted by the first author and lasted from 25 to 50 minutes, with an average of 35 minutes.

All interviews were transcribed with an automated transcription service. The first author read through the transcripts, noting remote work scenarios, working strategies, and challenges. The first and second authors discussed emerging themes and responses and refined the data into four parts outlined in the findings.

\subsection{Phase 2: Hypothetical Assistant Study}
Following the interview, the first author introduced our prototype via a pre-recorded demo video (see design details in the Phase 2 section). Ideally, participants would have been co-located with the research team to interact with the system themselves; however, this was not possible due to restrictions in our area. The video showed a research team member operating the prototype system to convey the idea to participants \sys{} (Figure~\ref{fig:robot}). Prior work has found it useful to use a prototype as a prop to stimulate discussion on technologies that may be unfamiliar to participants. ~\cite{wang2016flying, denning2014situ}. After the demo, participants were able to ask questions about the probe's design and operation.

\begin{wrapfigure}{r}{0.5\textwidth}
\centering
\includegraphics[width=0.5\textwidth]{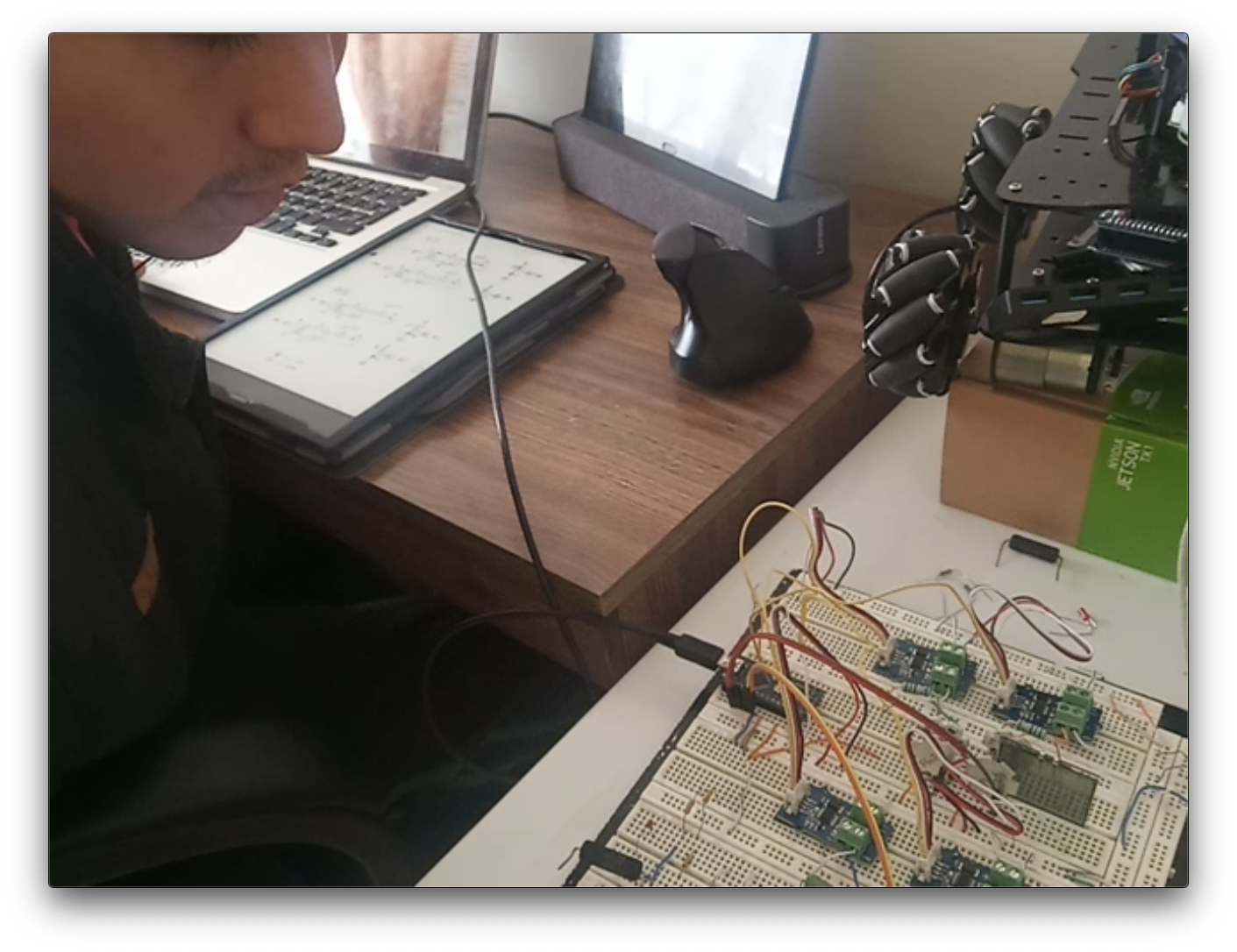}
\caption{A screenshot of a participant verbalizing a request to the hypothetical assistant while working on the hardware aspect of their embedded systems project during the study in Phase 2.}~\label{fig:workspace}
\end{wrapfigure}

Following the demo, we asked participants to enact a hypothetical work scenario with an experienced remote assistant. This approach is inspired by user enactment studies, which "investigate radical alterations to technologies’ roles, forms, and behaviors in uncharted design spaces"\cite{odom2012enactment}. The developers worked on their own embedded systems projects for at least 20 minutes, speaking aloud as if to a remote helper anytime they desired support (Fig.~\ref{fig:workspace}). We also asked participants not to limit requests based on the current prototype's perceived capabilities. This allowed us to better understand the full range of assistance desired, if there were no limitations.

To capture the work and requests, we asked participants to video-record the scenario enactment, in addition to the researcher's observations. This recording provided us spatial context information about their workspace and behavior for our analysis. After the hypothetical assistant study, we conducted a short debrief with participants to understand the nature of their project and to gather additional design suggestions the enactment inspired. The audio-recordings from the enactment and debrief were transcribed and analyzed alongside the Phase 1 interview data. The video-recordings were used to revisit and verify the researcher's notes and observations during the study.

\begin{table}[]

     \resizebox{\columnwidth}{!}{%
  \begin{tabular}{ccccccc}
    \toprule
    PID & Gender & Age & Occupation & Programming skill & Electronics skill & Collaboration experience \\
     \midrule
    P1 & M & 31 to 36 & P & 7 & 6 & 6 \\
    P2 & M & 24 to 30 & P & 6 & 6 & 5 \\
    P3 & F & 24 to 30 & P & 6 & 6 & 7 \\
    P4 & F & 24 to 30 & P & 5 & 6 & 5 \\
    P5 & M & 24 to 30 & P & 5 & 6 & 6 \\
    P6 & M & 24 to 30 & P & 5 & 5 & 4 \\
    P7 & M & 24 to 30 & P & 6 & 7 & 4 \\
    P8 & F & 24 to 30 & P & 5 & 5 & 6 \\
    P9 & M & 24 to 30 & S & 6 & 5 & 4 \\
    P10 & M & 24 to 30 & S & 6 & 5 & 7 \\
    P11 & M & 18 to 23 & P & 6 & 7 & 5 \\
    P12 & M & 31 to 40 & P & 4 & 4 & 2 \\
    \bottomrule
  \end{tabular}
    }
    \caption{Participants' background information. For the Participant ID (PID) column, P1-P12 represent the 12 embedded systems developers. For the Gender column, M = male and F = female. For the occupation column, P = professional developers  (paid to develop embedded systems projects) and S = student developers. Programming skill, electronics skill, and collaboration experience are on a 7-point Likert scale where 7 means expert and 1 means no experience.}
\label{tab:study3_stat}
\end{table}

\subsection{Study Limitations}

Our study is a first step towards a deeper understanding of remote assistance for embedded system development. Although promising, our study has a few limitations. Due to the remote nature of the study, we were unable to allow participants to interact directly with the study prototype. Hands-on interactions may have yielded a richer enactment with even more design suggestions and focused feedback. Second, the hypothetical assistant gave no response to participants' requests. Although this setup allowed us to envision the type of requests that were made, the behavior patterns we observed only reflect the first stage of help seeking: asking questions. Hands-on support would likely require significant interaction between the remote assistant and the help-seeker, which we did not capture in this study. Finally, we did not explicitly ask about or control for participants' prior experience with robot arms. Therefore, their prior experience might have affected their opinions and interactions in the hypothetical scenario, potentially limiting the range of requests participants believed possible.

\section{Phase 1 - Understanding Remote Work Scenarios and Challenges}

From participants' accounts of co-working scenarios, we identified distinct methods of working together. Question-asking (Q\&A) is where a developer had a question or needed information from external resources, including documentation, community forums, and direct peer-to-peer inquiry. General Support involved task-oriented requests that required expertise and access to the developers' workspace. Collaboration entailed two or more developers sharing tasks while working concurrently on a project. Q\&A and Support, which we refer to together as remote assistance, could be sought from known experts or from a crowd, while collaboration typically occurred with a team member.

\subsection{Remote Collaboration and Remote Assistance}
Participants discussed working collaboratively with a development team, working in a developer-client context (where the developer was hired to create a system or solve a problem with a client's existing system), and working individually while periodically seeking help and support from peers and more experienced developers.

Participants working with a partner or team discussed two ways of working on a project together. The first method was "Duplicate and Distribute," in which a team or pair duplicated all their hardware so that each member had exactly the same setup to work with. This method was useful for developer-client scenarios where the developer needed to create the system first before guiding the client in recreating the setup. The second method was "Divide and Conquer," where teams would distribute tasks to avoid the need for direct collaboration among team members. Since true distributed collaboration is difficult, this method was useful for remote teams to manage coordination costs.

In addition to working on a common project, participants periodically sought support on specific aspects of their work. These are the scenarios of \textbf{remote assistance} that we focus on in this paper. In these cases, the helper might not have the same hardware and software setup, so the developer might need to share project details to get effective help. The support sessions participants described were usually one-on-one, with participants primarily relying on experts that they know personally, including graduate advisors (P9, P10), colleagues (P1, P3, P4, P7), and friends (P5, P8, P9-P12). They used a combination of communication tools, including Zoom, Skype, Slack, and email.

Participants also described seeking help from others via community Q\&A sites (e.g., StackOverflow), developer forums (e.g., Arduino forum), and chat rooms (Discord servers). In these cases, participants described difficulties finding updated information or details specific to their hardware configuration. While most hardware components have specific documentation, hardware configurations were often unique to the project, and software interfaces were not always provided. In addition, embedded systems posts often include multimedia (such as images, a PCB design file, or video demos) which made it difficult to use text-based search engines to locate relevant information.
\begin{quote}
   \textit{There are several versions of machines or hardware like ESP8266...I might not know which version I'm using, but from this Instructables website, I can see that they're using this version so I might buy that version and try to use that, and when [I] implement that one, it does not work because the firmware and the hardware is totally different. (P6)}
\end{quote}

\subsection{When Is Remote Assistance Sought?}

\subsubsection{Finding Exact Answers}
All 12 participants reported relying on existing documentation, tutorial blog posts, or existing answers provided on Q\&A sites. While a substantial amount of this information exists, participants pointed out a crucial issue: they can often only find information containing answers that are similar to what they need instead of an exact match.

\begin{quote}
    \textit{There are so many similar solutions that exist on websites or the internet, but you have to choose the right one for your project. You have so many variables---you have similar MCU settings, but a little difference in between them. (P3)}
\end{quote}

The mismatch may be due to differing hardware configurations, outdated documents and information, or varying levels of detail provided across developer support communities. To address the information gap between similar or outdated project details, developers would resort to trial and error to determine a solution. While this type of iterative problem solving is useful in educational contexts, it is time-consuming and expensive in professional contexts. In these cases, experienced developers prefer to ask someone knowledgeable about their device or system rather than trying to find the information themselves. For example, P9 considered information searching to be a distraction from their main project:

\begin{quote}
   \textit{I skip the knowledge seeking and all [of] that contextual information. I skipped that step because I know that it's largely going to be a waste of my time when I know that there's someone who knows the answer already. Especially in the type of work that I do, a lot of this knowledge is like a one-time-use knowledge, like I need it for this specific project. (P9)}
\end{quote}

\subsubsection{Getting Immediate Help}
Eight participants discussed the need for quick and timely feedback when working on embedded systems projects. While the desire for on-demand help in programming has been well-documented~\cite{chen2016towards}, embedded systems developers' needs may be exacerbated by the higher context-switching costs on complex tasks, which can mean that synchronous help sessions can often be more beneficial. For example, P1 noted that asking for remote help might result in interruptions for both parties:
\begin{quote}
    \textit{You can ping someone, and...they might be working on like, different projects, you know, and so they don't want to do the mental task switching. So they just ignore that message until they start working on that project again, and by that time, you might not be working on that project. (P1)}
\end{quote}

P1 mentioned that in person, "I can just turn around and see" if asking for help would interrupt a potential helper or not. Given the coordination difficulties, participants preferred synchronous communication for immediate help over delays.

\subsection{Tasks and Strategies for Remote Assistance}
In this section, we report strategies participants described to enable remote question-asking, assistance, and collaboration. We also note challenges developers face when seeking remote support in system development. To enable a minimal level of remote collaboration with embedded systems, participants have used strategies like taking pictures of their circuits, live-streaming during a conference call, marking up images to explain their work, and writing explanations to post in forums. While these practices are similar to those of software development, they were not as effective in hardware development.

\subsubsection{Sharing Visual Context}

When asking questions remotely, participants expressed a need for better annotation tools to provide visual context. Three participants mentioned that they often include visual information to communicate their problems and system setup on forums or in chatrooms. For example, P2 describes asking for help over Slack and exchanging annotated images with a mentor:

\begin{quote}
    Right now my advisor and I, we had a discussion over Slack over debugging. And basically what ended up happening is like, I drew the diagram on my iPad and then I sent him a picture of the circuit and then I use colors to draw where I'm plugging in the scope. And then he sent me like, a marked-up image back. (P2)
\end{quote}

However, help-seekers and remote helpers face difficulties providing enough visual detail to effectively communicate about the devices. While simple directions were possible, as in the scenario above, many cases also found that the complexity of the circuits made providing the necessary visual context inconvenient. Although advanced circuit drawing tools provided more functionality, they were viewed as too inaccessible for the types of tasks our participants engaged in remotely, like help-seeking and brainstorming.

\begin{quote}
   A lot of the annoying things that we're doing is like, for instance, if we were physically together in person, I'd be able to draw the circuit. But if it's a more complicated circuit, it's a lot harder to like send detailed drawings of the circuit. So...we're using professional software like Altium, a PCB layout software, to draw a circuit, because that's the way that we're facilitating remote collaboration right now. But in the real world, like nobody would do this. Like nobody would brainstorm using a professional tool. (P9)
\end{quote}

Despite having access to a range of drawing tools, participants needed flexibility to choose a tool at the right fidelity.
In addition to the challenges of custom drawings, taking photos was also difficult. For example, P11 tried taking photos to augment their Q\&A, but found it hard to capture the entire board:

\begin{quote}
    I have another board that's more---from a circuit point of view---complicated. And you can try taking pictures of it. But when wires overlap, it's really hard to read a picture, because you're often like rotating a board to see different angles of it. (P11)
\end{quote}

These visual challenges were not limited to asynchronous collaboration. When asked about techniques that would support help-seeking in synchronous sessions, half of the participants (6/12) mentioned a desire for better viewing tools so that "my collaborators can see where I am pointing at [and] looking at when speaking" (P4). Four participants noted that they had to provide additional visual cues during remote collaboration when compared to in-person communication, including "gestures" (P1, P6) and "words" (P9, P10) to compensate for the insufficient context caused by the limited viewing angle of their camera.

\begin{quote}
    I would want something where people can see what I'm looking at. And they can see the way that the wires are connected to each other. (P6)
\end{quote}

Participants wanted to point a camera at the physical components to highlight specific issues as they spoke. However, those using webcams struggled to capture and share the details of their setup while discussing their problems.

\begin{quote}
    What I have to do is, I have a Zoom on my laptop, and I rotate the laptop so that [the helper] can see what I'm doing on the screen while I'm modifying the hardware. (P7)
\end{quote}

The need for better visual context was so significant that some participants opted not to use synchronous remote communication tools like Zoom or FaceTime because it was more difficult to explain a project compared to sharing pictures and pre-recorded videos. For example, P9 preferred Slack over video chat because of the camera focus:

\begin{quote}
    I don't think like a video chat is that valuable because the video chats focus on face-to-face communications. Whereas with Slack, I can send a picture of my circuit, I can send a picture of my oscilloscope output...I can send the code of my circuit, the code snippet. So video chatting doesn't necessarily facilitate the important parts of debugging. (P9)
\end{quote}

These examples highlight that attempting to use communication tools intended for other purposes, like face-to-face video chatting, is insufficient for communicating about embedded systems work. Participants seeking remote assistance desire a rich visual communication channel to show their embedded system hardware setup as well as their eye gaze and gestures.

\subsubsection{Physical Direction and Manipulation}
Hardware configuration, such as the placement of wires and connecting components, is a prime opportunity for mistakes and problems to arise, yet difficult for requesters to capture. Hardware solutions, such as a response to a help request, are similarly difficult for helpers to demonstrate. As P4 explained, hardware modifications were challenging because of the interplay between code, connections, and the physical position of the hardware:

\begin{quote}
    Because in circuit work it depends a lot [on] connections...it is not just copy and pasting a coding block...Sometimes, [requesters] make the incorrect connection [on] the incorrect components usually. How you place the components is also important; maybe they [ruined the hardware]. So, when working with hardware, things are very sensitive. (P4)
\end{quote}

Unlike software development, a solution to a hardware problem would not be a simple "block of code" to fit into one's existing work. Sometimes the identified problem is in the setup of the components, which requires tacit knowledge that might not be included explicitly in the solution.

Many participants further reported that their setup—with multiple components spread out across a workbench—felt cumbersome to explain in any context. Five participants reported that sharing the necessary visual context often required a different setup than the one they used for working: they might have to rearrange components for a good camera angle, or they might have to break down components to expose the elements that they needed help with. For example, P7 mentioned the difficulty of finding a good camera position to share their setup synchronously.

\begin{quote}
   I was in a Zoom meeting, and I was using the laptop again, and I wanted to show something very small and specific in the board, and I wanted to use like, an external camera to, to transmit to Zoom, and I couldn't find a way to do that. And sometimes it is a small circuit board—you can’t really lift the circuit board because there's a lot of things attached to it. (P7)
\end{quote}

These difficulties span Q\&A, support, and collaboration scenarios, as effective communication about a hardware setup, problems, and proposed solutions is necessary in each context.

Four participants mentioned that they wished the remote helper could simply operate on their devices to save them the redundant effort of carrying out a described task. For example, P8 described how difficult carrying out a simple "push this button" instruction from a remote helper could be, compared to doing it directly:

\begin{quote}
   Managing the hardware [is difficult], even if it's something as simple as like, I need to press a button...Oh, which wire? Oh, the red one. Which red one? [There] are three. Oh, the one on the bottom? Like, it's—it's a lot more difficult. And when you're in person, they can just move that wire, right? (P8)
\end{quote}

This suggestion to allow remote manipulation is beyond what prior work has explored, as existing research has proposed tools for better guidance rather than direct manipulation of physical devices.

\subsection{Summary of Interview Findings}
Remote assistance in embedded systems work ranges from question-asking to task support to collaboration on a shared project. Participants described appropriating synchronous and asynchronous communication tools to work with peers, mentors, and clients. Common forms of remote assistance were to get specific information difficult to find in online archives, to request help executing or debugging specific tasks, or to provide guidance integrating a hardware solution into a project. Participants desired richer communication channels for sharing hands-on visual context remotely, including their workbench, system setup, and non-verbal communication cues, along with existing audiovisual capabilities. Currently, these collaboration scenarios also require significant manual interaction, which, when possible, can be physically uncomfortable or require a developer to rearrange their entire workspace. The need for physical manipulation support motivated the design of Handy, the technology probe described in the next section. With this as a prop, participants delved more into what future remote collaboration on embedded systems could look like, moving beyond existing tools and capabilities.

\section{Study Phase 2}
For the second phase of the study, we aim to explore the possibilities for enhancing remote assistance for embedded systems developers.

\subsection{Probe Design Rationale}
We designed and demonstrated "Handy," a web-based programming support tool with an augmented, teleoperated robot arm that allows remote collaborators (helpers) to perform inspecting, wiring, and coding tasks upon the end-user developer's (requester's) request~\cite{chen2020wireon}. Handy provides both communication support and direct helper access to an embedded systems project. Its design is inspired by the Heimdall system~\cite{karchemsky2019heimdall} which allows a remote course instructor to visually inspect a student's prototype circuit by taking pictures via a mounted fixed-position camera and provide verbal guidance and direction. Heimdall also provides a remotely accessible wiring system, allowing the instructor to "inject" signals and test wiring.

\begin{figure*}[b]
\minipage[t]{0.51\textwidth}
  \includegraphics[width=\linewidth]{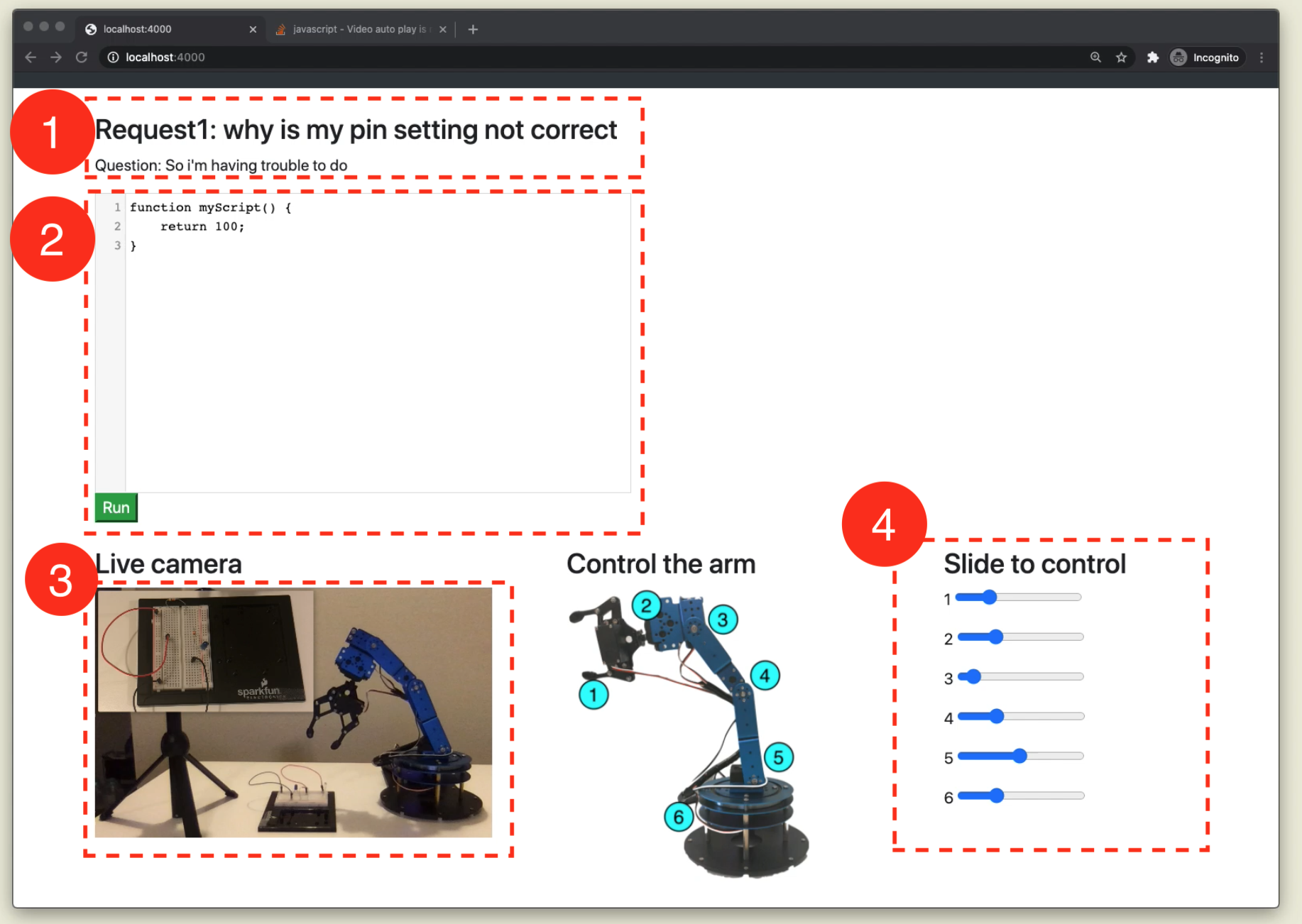}
  \caption{Web UI controls. This includes four parts: 1) a natural-language task description, 2) a code editor that includes the task-relevant code, 3) two fixed-position live stream camera views, and 4) a six-slider control panel that corresponds to the six servos of the robot arm.}~\label{fig:web}
  \Description[Fully described in the text.]{Fully described in the text.}
\endminipage\hfill
\minipage[t]{0.44\textwidth}
  \includegraphics[width=\linewidth]{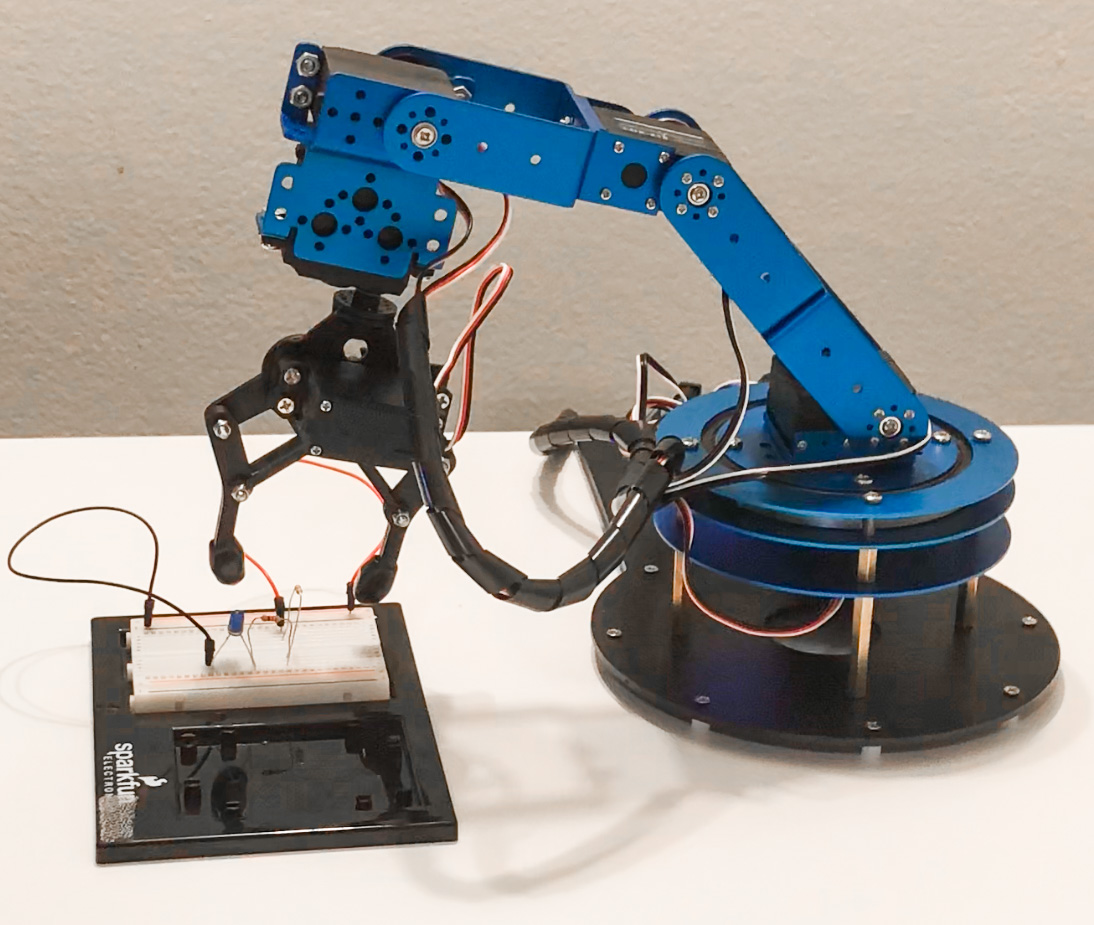}
    \caption{A photo of the robot arm performing a wiring task. }~\label{fig:robot}
    \Description[Fully described in the text.]{Fully described in the text.}
\endminipage
\end{figure*}

Handy builds on the success of Heimdall's features to facilitate collaboration beyond student instruction. Our system elicits scenarios common for professional developers, such as the helper making circuit changes independently of the developer, the developer handing off a task to the helper while working on something else, or the developer and helper working together on an open-ended problem. The system consists of two parts: a robot arm (Fig.~\ref{fig:robot}) and a web user interface (Fig.~\ref{fig:web}).

\textbf{Robot Arm:} To enable physical manipulation, we used an off-the-shelf robot arm (USD \$59.00)~\cite{hiwonder} with six degrees of freedom that could perform simple pick-and-place tasks. The six servos are connected to an Arduino Uno board, which is controlled by a web UI.

\textbf{Web User Interface:} To enable remote control, we designed a web UI that consists of four parts (Fig.~\ref{fig:web}): 
1. A natural-language task description that specifies the end-user developer's request; 
2. A code editor that includes the task-relevant code; 
3. Two fixed-position live-stream camera views, including one top-down view and one side view; 
4. A six-slider control panel that corresponds to the six servos of the robot arm.
We used two Logitech C270 cameras that capture 1280x720 resolution in live-viewing mode. We adopted a fixed-position camera view rather than a moving and remote-controllable camera view, as prior studies have shown that some users have privacy concerns surrounding the latter, specifically drones ~\cite{10.1145/3025453.3025907, wang2016flying} and teleoperated robot arms~\cite{10.1145/2696454.2696484}. We decided that a fixed-camera view would minimize privacy concerns without sacrificing usability.

\section{Results}

First, we synthesize support needs of remote embedded systems developers by analyzing requests made to a hypothetical assistant. 
Second, we present concerns that embedded systems developers have regarding engaging with crowd help or an automated assistant.

\subsection{Requests Made of the Hypothetical Remote Assistant}

During the hypothetical assistant elicitation study, we asked participants to say the word "agent" before they made a request to make it clear when they were asking for support from the hypothetical assistant. Overall, each participant made 25 requests on average (max: 54, min: 8). Participants often phrased requests in a way that required knowledge of their hardware (e.g., circuits), the specific context (e.g., where they were pointing), and their code base. We found that only 12\% of participants’ requests were "self-contained" and needed no context to understand (e.g., "What is the difference between VDD and ground?").

The following section details the types of requests participants made of the hypothetical assistant and shows the range of assistance developers desire when working on projects. The following list outlines the six request categories that we coded from "Agent: <request>" utterances, the number of sessions in which each request type appeared, and the overall frequency of each request type.

\textbf{1. Physical Aids}
All participants made \textit{physical aid} requests, asking the agent to either perform an action for them, such as plugging in a USB, or to provide assistance on tasks they could not perform alone.
\begin{quote}
    \usersaid{Can you hold the circuit board tight so I can solder something on it? (P7)}
\end{quote}
\begin{quote}
    \usersaid{Please connect my Arduino to my computer. (P10)} 
\end{quote}
These operation requests are simple "helping hands" tasks that do not require special knowledge but help the developer complete a physical hardware task more quickly or efficiently. In follow-up interviews, a participant explained that this kind of help would save time dealing with projects that require repetitive actions.

\textbf{2. Measurement Aids}
All the participants asked the agent to measure and report signal output, such as the current of their circuits, and to test data connections. For example,
\begin{quote}
 \usersaid{Can you check the borders in GPS and check their serial data? (P1)}
\end{quote}
\begin{quote}
  \usersaid{Is this device sending data to the server? Or is there any problem with this data-sending issue? (P1)}
\end{quote}
These measurement aid requests reflect common hardware debugging tasks that provide essential low-level information to embedded systems developers about the current state of their system.

\textbf{3. Wire Refactoring}
Three participants also made "wire refactoring" requests to have the agent rearrange their circuits and components. As with code refactoring, these participants wanted a cleaner and easier way to find or measure certain components of their circuits.
\begin{quote}
\usersaid{Can you clean up this voltage divider? (P9)}
\end{quote}
\begin{quote}
\usersaid{Could you clean like, my circuits? They are like, super messy right now. (P5)}
\end{quote}
Tedious but necessary, wire refactoring can be a cleanup step to reclaim material (i.e., excess wires), or a step to prepare for the next stage of prototyping, such as developing a printed circuit board.

\textbf{4. Debugging and Bug Fixing}
Most of the participants (10/12) made debugging requests. In embedded systems, software errors can be caused by faulty code or by an incorrect hardware setup.
\begin{quote}
    \usersaid{Why am I getting this error? (P9)}
\end{quote}
\begin{quote}
   \usersaid{I did not hit the reset button, but I still see the device in the drop-down menu, and the memory layout is blank. Why is that the case? (P11)}  
\end{quote}
Hardware debugging requires specialized knowledge about how the circuit, component, or code should be set up, as well as semantic details about how errors might occur.

\textbf{5. Explanation Requests}
In addition to debugging help, close to half of the participants (5/12) sought explanatory support. They asked the agent to help them understand things relevant to their projects.
\begin{quote}
    \usersaid{What does 'Nordic DFU trigger interface was not found' mean? (P1)}
\end{quote}
\begin{quote}
    \usersaid{What is the difference between VDD and ground? (P6)} 
\end{quote}

\textbf{6. Reference Aids}
More than two-thirds (8/12) of the participants asked the assistant to find a component datasheet or to recall information from relevant device documentation and programming APIs.
\begin{quote}
    \usersaid{For this device, how much force is concealed during operation? (P11)}
\end{quote}
\begin{quote}
    \usersaid{Pull up the datasheet for the FPGA; I'm trying to figure out...how much current this microphone is supposed to take, because it's taking up more current than I was originally expecting. (P5)}  
\end{quote}

\textbf{7. Strategic Guidance}
Half of the participants (6/12) also got stuck at some point during their development, in which case they requested high-level hints or guidance from the assistant.
\begin{quote}
    \usersaid{How do I delete a characteristic that I added to generic access instead of the battery service? (P7)}
\end{quote}  
\begin{quote}
    \usersaid{I am trying to hit the delete button, but I am not getting the value. What else should I do? (P11)}
\end{quote}

Guidance requests were specific to the project and task, and contextual knowledge was required to form a proper answer. High-level direction could also yield "opinionated" solutions, reflecting best practices and developer preferences for a particular approach to a problem.

\subsection{Concerns About On-Demand Help Solutions: Invasive, Unsupervised Interaction}

As we demonstrated, participants engaged with a remote collaborator in various ways and envisioned a range of uses for an assistant. However, there were three main concerns expressed in relation to remote or automated support tools for embedded systems developers, including privacy, physical control of the robot and its potential risk, and trust in the remote helpers.

\subsubsection{Privacy}
The most cited concern among participants was maintaining privacy due to the movable camera and speaker (10/12). This concern is less pronounced if it involves a stationary camera (e.g., always facing one part of the workbench) and no audio input is captured (e.g., text communication). Participants worried about their project ideas being stolen (P1, P4), their conversations with others in the room being overheard (P1, P6, P11), or their face or room being seen (P4, P11). These concerns were raised in response to the idea of a live video feed directed into the workspace of a helper, and the possibility that the direction of the camera could be remotely controlled by a helper.

P2 explained that they felt uncomfortable with one-way anonymity, especially when exposing potentially identifying information over video chat or a narrated video.
\begin{quote}
     \usersaid{So the other person on the other end is anonymous, right? I don't know who they are. But I'm not anonymized to them... They have all of this identifiable information. They know what I'm working on. They can hear my voice. (P2)}
\end{quote}

This suggests that participants have privacy concerns about their projects as well as their identity. Concerns about privacy also extended to unintentionally invading the privacy of in-person co-workers in close quarters when engaging with remote helpers using a live feed. Participants who did not have these concerns mentioned that they have their own workspace which contains little personally relevant information.

To mitigate these concerns, participants suggested integrating visual privacy-protecting techniques into the system, such as "only have the device that I am pointing to visible, and blur the outside view" (P4). Prior work has studied users' perception regarding different blurring methods applied to a movable robot arm and suggested that using an image depth filter (i.e., objects extremely close to the camera appear black, whereas farther away objects are lighter) is perceived to best protect privacy~\cite{klow2019privacy}.

\subsubsection{Physical Control and Risk}
More than half of the participants (6/12) were also concerned about the precision and flexibility of the robot arm, expressing that they would need to first know how capable the arm was before using it on critical tasks. Three other participants worried that the robot arm could damage their devices (P8, P9) or introduce interference into their circuit (P6). 
\begin{quote}
    \usersaid{You're using a microphone sensor, and you're trying to check for [a] certain frequency of something. And you would like the robot arm to place a probe at a point...to check the frequency. I wonder if the robot is going to cause interference [in] the project, like sound frequency or like mechanical effects on the performance. (P6)}
\end{quote}
This statement further emphasizes the importance of risk control for an agent engaging in physical support work.
Allowing physical interaction with a project carries additional risks. A sensor component knocked to the side or wires crossed with live current could destroy devices or setups that are expensive or hard to replace. Participants' concerns reflect the current lack of fine precision in amateur off-the-shelf physical manipulation tools, as well as uncertainty about how well a helper can remotely control a device. However, even simple manipulators could be improved enough to correspond to the level of detail typical of embedded systems projects. For example, prior work has also studied the performance of various control settings when humans maneuver a robot arm. They suggested that for users with no background in robot control, allowing the robot to have some autonomy over the arm can help reduce error rates~\cite{leeper2012strategies}.

\subsubsection{Trust}
When considering the possibility of a remote helper that could operate a robot arm to work asynchronously (i.e., when the help-seeker is not present), nearly half of the participants (5/12) indicated hesitation with allowing unsupervised access to their projects and potentially their entire work and living space. While the prospect of an autonomous assistant was seen as "useful" and a "time-saver," participants suggested placing limits on its capabilities. For example, P9 and P10 wanted some assurance to prevent the robot arm from "damaging" their artifacts.

Ensuring that the system was fully deactivated would be one way of guaranteeing that the device could not be used for unintended purposes without users' knowledge. Aside from a mistrust of potential bad actors snooping around, three participants had concerns over whether they could trust the robot assistant and its solutions. For example, P8 referenced the "intelligence" of the assistant, indicating that they expected some decision-making ability when helping on a task.

\begin{quote}
    \usersaid{It has to be very sophisticated to work with these [types] of complex circuits, and it has to be very intelligent for solving [these types] of uncommon issues or this type of issue as a firmware development software. (P8)}
\end{quote}

These concerns indicate that participants wanted more assurances about both the capabilities of the automated assistant and the expertise and trustworthiness of the helpers up front. Future systems could consider incorporating a "demonstration mode" to highlight operation modes, limitations, and capabilities to make it clear what to expect of an embodied autonomous and/or remote assistant.

\subsection{Summary of Findings}
Participants envisioned a range of potential uses for automated and semi-automated physical manipulation tools, from a smarter helping-hands tool to an interrogable debugger. With clear need and support expressed for remotely accessible physical manipulation tools, participants also described their concerns with the privacy, controllability, and trustworthiness of these types of systems. In the discussion, we reflect on these findings and offer suggestions for future work in remote collaboration tools.

\section{Discussion and System Design Suggestions}
This study investigated the nature of remote collaboration and assistance for embedded systems developers and the potential for future tools and platforms to support these scenarios. Our findings identified two primary technical shortcomings that make remote collaboration inefficient and undesirable with current collaboration tools: lack of support for visual inspection and lack of support for direct manipulation. Inspection needs include communicating the physical setup, sharing visual context during communication, and arranging a workspace to allow for remote access. The direct manipulation needs include physical handling of hardware and associated components, both during work and during communication, and enabling electrical and digital remote control.

The findings from our enactment study with a hypothetical assistant paint a rich picture of future work styles enabled by a remote helper with physical capability. We observed seven types of assistance requests during enactments, from providing guidance and explanations to refactoring wires and checking measurements. Although our scenario focused on human help, participants also enacted scenarios that with tasks that an automated or semi-automated agent could perform on their behalf. In the following sections, we reflect on the challenges and support needs identified in the study, taking note of solvable challenges, near-future solutions, and areas in need of further research.

\subsection{Visual Tools with Privacy-Sharing Controls}
To help requesters more easily set up and share their physical view and gestures, we suggest using a teleoperated, 360-degree camera that provides a wide-angle view. Our findings reveal the need for better visual inspection support. Developers could not use embedded laptop webcams, optimized for video chatting, to effectively record or livestream images of their project. During remote conversations, participants needed several cameras to show their face and gestures along with their workspace and physical components. A multi-angle view might also enhance recorded video for asynchronous support requests, for example, allowing developers to easily juxtapose an overview to provide context and tight shot to focus on a problem area. Yet, even participants who used an additional camera to provide a better view of their work, such as an external webcam or a phone camera, needed the ability to reposition the view to focus on different aspects of their work during a conversation or recording session. A dynamic camera angle would also be useful to remote assistants, allowing them to pan and zoom as necessary to get a better understanding of the project they were helping out with.

Although viewability could be enhanced via multiple, dynamic camera angles, participants were concerned about unwanted visual intrusion beyond the scope of the workspace. Fortunately, a well-designed viewing platform could constrain the field of view to minimize the possibility of wandering eyes, while maximizing the utility of a multi-angle setup. Prior work exploring the privacy-utility trade-off for remotely teleoperated robots (e.g., iRobot) found that blurring the peripheral view using a color-skewed superpixel filter provided the same visual utility to operators as no filter, while offering more visual privacy to those co-located with the robot. In addition, the system could adopt the target-tracking technique proposed by Rakita et al., to force a movable camera to stick to a pre-defined viewing area while adjusting its location and zoom. As these features preserve visual privacy, they also would make it easier to efficiently update the view, whether via local interface or web-enabled remote interface.

However, these constraints do not fully address issues of trust that participants raised when considering the possibility of allowing a remote assistant to access a workspace and perform tasks independently. We will return to this concern in a future section.

\subsection{Annotation Tools for Contextual Explanation}
A second issue revealed in our findings is the need for richer communication channels with annotation support. This need was present in both synchronous and asynchronous scenarios, where participants struggled to adequately explain their system, setup, and problem using only verbal explanation and captured images. Developers may benefit from recent advancements in augmented visual streams, which allow people to augment real-world imagery with virtual information.

Augmented Reality (AR) technology has applications in many domains, including planning, gaming, and education, but one of the most often cited use cases for AR is remote guidance, where a remote helper assists a worker in performing a physical task. Remote guidance is valuable in a variety of situations, particularly for tasks in which workers need to reference helpers' expertise. AR has the potential to improve communication in remote guidance tasks by allowing helpers to augment a worker's physical space, adding instructions to help them perform a given task.

Previous literature has proposed two primary ways for helpers to augment a worker's physical space to give remote guidance: annotations and telepresence. Annotations are virtual notes or marks that remote helpers leave in a worker's space as guidance. Telepresence represents all or part of the helper's body virtually and displays that representation in the worker's space, allowing them to communicate through gestures and other non-verbal communication.

In addition to dynamic viewing angles discussed in the previous section, object-tracking annotation tools allow developers to create virtual notes or marks, allowing them to sketch 3D annotations, highlight objects, or provide real-time guidance during hands-on work. An AR system could even be configured to record a digital twin of a project's current configuration. This could enable developers to demonstrate a physical change before making it on the actual hardware or even be used as version tracking to better facilitate trial-and-error problem solving.

The addition of annotation and telepresence tools might help to address concerns about unequal disclosure in remote assistance scenarios. Interacting with strangers online comes with risk, however, a channel that allows a developer to interact with others via an avatar might provide enough sense of anonymity to assuage these concerns.

\subsection{Providing Hands-on Assistance}
In our hypothetical assistant enactment study (Section 5.3.7), four out of seven request types necessitated that an assistant have direct physical interaction with the system. These included providing physical aids, taking measurements, refactoring wiring, and assisting with debugging and bug fixing, which are currently not feasible with the tools and platforms available. However, research is progressing in this area. Tools like Proxino~\cite{wu2019proxino}, VirtualComponent~\cite{lee2021virtualwire}, and CircuitSense~\cite{wu2017circuitsense} have begun to bridge the gap between software and hardware, simplifying the process of visualizing, sharing, and manipulating hardware configurations.

To further advance this field, we suggest equipping the entire workspace, rather than just individual circuit boards. We advocate for the use of a robotic arm, as demonstrated in our study, because it offers greater scalability for expert development that typically involves various hardware components, each requiring different configurations and setups. Moreover, a versatile manipulation tool could support additional tasks such as holding components during testing or performing pick-and-place operations.

Despite the potential benefits, many participants expressed concerns about the quality of assistance from a robotic helper, citing issues with precision and flexibility. While we anticipate that the capabilities of robotic devices will improve, there is apprehension that helpers might still struggle with maneuverability and lack of experience in controlling robotic arms. To enhance remote teleoperation, we propose two near-term solutions: integrating autonomous functionalities and facilitating routine task automation.

Studies have demonstrated that incorporating autonomous movements and planning into a robot arm can significantly aid humans in managing complex tasks. For instance, Leeper et al. found that higher levels of autonomy in a robot arm led to better task performance by users~\cite{leeper2012strategies}. Similarly, Kent et al. showed that a point-and-click interface, which automatically suggests grasp positions, can help users complete tasks more efficiently and with fewer errors~\cite{kent2020leveraging}. Future systems could include a degree of autonomy in their robot arms to simplify operations for remote helpers.

Furthermore, many tasks that participants needed help with are repetitive. Systems, like Mimic~\cite{mahadevan2022mimic}, uArm~\cite{xarm} or Universal Robotics' cobots~\cite{universal}, are already automating such processes both in research setups and industrial environments. More research to enable robots to learn from human demonstrations through machine learning could extend these capabilities to more complex and varied tasks not yet widely adopted~\cite{grollman2010incremental}.

In contexts involving embedded systems, not every support request necessitates human intervention. Tasks such as providing limited physical assistance, taking measurements, and wire refactoring could potentially be handled by an intelligent agent. Future support systems might also incorporate additional testing functionalities, like probes to inject and measure signals and simple displays to indicate outputs, thereby allowing developers to delegate more tasks to the robot arm, enhancing productivity and easing the load on remote human helpers.

\subsection{Trust and Social Norms}

While technical enhancements can address certain remote collaboration problems, many issues may exist outside the scope of tools and technologies alone. Interpersonal trust is crucial for cooperation in large organizations, especially when long-term relationships have not been established. However, this type of trust can be difficult to foster with online crowds where requesters do not always know helpers' identities or levels of expertise. To address these issues, we suggest:

\textbf{Version Control for Physical Computing}: Implementing version control systems similar to those used in software development, such as Github, can provide developers control over the changes made by remote helpers. Although versioning is challenging with hardware and changes causing catastrophic errors cannot be reversed, simulation tools and virtual solutions could prevent physical damage.

\textbf{Enhanced Communication Channels}: Providing visual representation of helpers and using avatars can help develop trust more effectively than text-based communication alone. Future systems could explore using video filters to reduce privacy concerns, allowing developers to engage more confidently with remote helpers.

\section{Conclusion}

This research explores the future of remote collaboration tools for embedded systems developers by adding physical task support absent in current communication and collaboration tools. Through a study with embedded systems developers, we identified the major shortcomings of current tools and conducted a user enactment study with a bespoke remote manipulation tool to elicit a range of requests developers would make of a collaborator. Our findings outline the scenarios and strategies for remote work, the challenges and support needs of developers, and their concerns about privacy, control, and trust when working with remote physical manipulation tools. This research contributes to the literature by aligning embedded systems development with remote, on-demand collaboration and help-seeking in software environments, providing a foundation for future work on remote manipulation tools and enhanced collaboration support.

\bibliographystyle{ACM-Reference-Format}
\bibliography{sample-base}

\end{document}